\documentclass[12pt]{iopart}
\usepackage{epsf}
\begin{document}
\jl{3}

\title{Longitudinal and transverse forces on a vortex in superfluid 
$^4$He}
 
\author{H M Cataldo, M A Desp\'osito,  and D M Jezek}

\address{
Departamento de F\'{\i}sica, Facultad de Ciencias Exactas y 
Naturales, \\
Universidad de Buenos Aires, RA-1428 Buenos Aires, Argentina\\
and Consejo Nacional de Investigaciones Cient\'{\i}ficas y
T\'ecnicas, Argentina}

\date{\today}

\begin{abstract}

We examine the transverse and longitudinal components of the
drag force upon a straight vortex line due to the scattering of liquid 
$^4$He excitations.  For this purpose, we consider a
recently proposed Hamiltonian that describes the dissipative
motion of a vortex, giving an explicit expression for  the
vortex-quasiparticle interaction.  The involved dissipative
coefficients are obtained in terms of the reservoir correlation 
function.
Most of our explicit calculations are
concerned to the range of temperatures below 0.4 K, at which the 
reservoir
is composed by phonon  quasiparticle
excitations. We also discuss some important
implications in the determination of possible scattering processes 
leading
to dissipation, according to the values of vortex mass found
in the literature.

\end{abstract}

\submitted
\maketitle

\nosections

Although it is well established that vortex dynamics plays an
important role in the behavior of superfluids, there are many
controversial results in the literature about its dissipative
motion. At zero temperature the motion of the vortex is provoked
by the transversal Magnus force, but at finite temperatures the
scattering of collective excitations by the vortex provides a
dissipation mechanism that damps its cyclotron motion.  It is
widely accepted that for low enough temperatures, and only
taking into account phonon-vortex scattering, the longitudinal
component of this dissipative force behaves as fifth power in
temperature \cite{ior}.  Nevertheless, the form of the
transverse component is still controversial
\cite{so} since very different results were reported so far and it does 
not appear that any of them can be regarded as definitive. 

In this context, we have recently presented \cite{ca1} a model for the
dissipation  of a straight vortex line in superfluid $^4$He  in
which the vortex is regarded as a macroscopic quantum particle
whose irreversible dynamics can be described in the frame of
Generalized Master Equations. This procedure enables us to cast
the effect of the coupling between vortex and heat bath as a
drag force with one reactive and one dissipative component,  in
agreement with phenomenological theories.  In this paper
we shall investigate the components of the drag force considering that
the reservoir is composed by the quasiparticle (qp) excitations of the
superfluid. The dissipative vortex dynamics arises then from the 
scattering
processes ocurring between the vortex and the qp's. Most of our 
explicit
calculations will be concerned to the range of temperatures below 
0.4 K,
at which the qp excitations constitute a phonon reservoir\cite{do}.

Let us begin by performing a description of our model.
Choosing a coordinate system fixed to the superfluid,  the
Hamiltonian  for a straight vortex line parallel to the $z$-axis, 
may be
written as \cite{ca1,do}
\begin{equation}
H_v = \frac{1}{2 m_v} \left [{\bf p} - q_v {\bf A}({\bf r})\right]^2 
\,\, ,
\label{Hfree}
\end{equation}
where  $ {\bf r} $ and $ {\bf p} $ denote respectively the vortex 
position
and momentum operators,
\begin{equation}
{\bf A}({\bf r}) = \frac{h\,\rho_s\,l}{2} (y,-x)
\end{equation}
is the vector potential for the Magnus force, $m_v$ is the  inertial 
mass of the vortex,  $\rho_s$  the number  density of the superfluid, 
$h$  Planck's constant, $l$ the system length along the $z-$ axis and
$q_v = \pm 1$  the sign of the vorticity according to the right handed 
convention. 

In what follows we shall consider the excitations to be at rest,
so that the velocity of the normal fluid vanishes.  In a
previous work \cite{ca2} we have considered an
interaction Hamiltonian of the form
\begin{equation}
H_{int} = - {\bf B}\cdot \,\, {\bf v}
\label{hin}
\end{equation}
where $ {\bf B} $ and $ {\bf v} $ are vectors that depend on the 
reservoir and 
vortex operators respectively. 
We have proven that the only linear combination of the vortex 
observables 
that leads to a dynamics  in accordance with  phenomenological 
descriptions 
is,
\begin{equation}
{\bf v} = \left(\frac{p_x}{m_v}- \frac{\Omega}{2} \,y, 
\frac{ p_y}{m_v} +
\frac{\Omega}{2}\,x\right)\,\,,
\label{vfree}
\end{equation}
being  $\Omega$ the ``cyclotron frequency",
\begin{equation}
\Omega= \frac{q_v \,h\,\rho_s\,l}{m_v}.
\label{Ome}
\end{equation}

In order to obtain an equation of motion for the mean value of the 
complex
vortex position operator $R = x + i y $, in previous works 
\cite{ca1,ca2}
we have derived by means
of a standard reduction-projection procedure, a generalized master
equation for the density operator of the vortex. We have employed a 
usual weak-coupling
approximation, in which the vortex dynamics is affected by the 
reservoir
degrees of freedom only through the second order time correlation 
tensor
$\langle {\bf B}(t) {\bf B}\rangle$, where the angular brackets 
indicate
an average over the reservoir equilibrium ensemble and ${\bf B}(t)$
denotes a free time evolution for the reservoir operators. 
In addition, such tensor is naturally assumed to be isotropic in the
$x-y$ plane. 
We have also made use of  the Markovian approximation which
assumes that such correlations are short
lived within observational times.
 From such  master equation, 
we then extracted equations of motion for the expectation values of 
vortex position and
momentum operators. Finally, after some algebra and by elimination of 
the momentum,
we obtained the desired equation for $\langle R(t)\rangle$ 
\cite{ca1,ca2}:
\begin{equation}
m_v\langle\ddot{R}\rangle
=i (m_v\Omega + \gamma) \,\langle\dot{R}\rangle \,\, .
\label{queda}
\end{equation}
The complex coefficient $\gamma$ is defined as 
\begin{equation}
\gamma = \frac{2\Omega}{\hbar}\int_{0}^{\infty}\,d\tau\, 
\psi(\tau)\,e^{\displaystyle - i\Omega \tau } \,\, ,
\label{integ}
\end{equation}
where $\psi(\tau)$ is the imaginary part of the time correlation 
isotropic tensor element,
\begin{equation}
\psi(\tau)={\cal I}m\,(<B_j(\tau)\,B_j>) \,\, .
\label{corr}
\end{equation}

In the right hand side of equation (\ref{queda}) one can
identify two forces, namely the Magnus force \cite{so}
and the drag force.
In particular, the drag force
\begin{equation}
F_D = i \gamma
 \,\,\langle\dot{R}\rangle
\end{equation}
has two components.  One is parallel to the Magnus Force 
 which we shall call the Transverse Force (TF) :

\begin{equation}
F_T = i\, {\cal R}e (\gamma)
 \,\,\langle\dot{R}\rangle,
\end{equation}
and the other one is parallel to the velocity
and we shall refer to it as the Longitudinal Force (LF) :

\begin{equation}
F_L = - {\cal I}m ( \gamma)
 \,\,\langle\dot{R}\rangle .
\label{flo}
\end{equation}

 We want to recall
that it is well known that the LF must  be opposite to
the velocity (in our notation this means ${\cal I}m (\gamma)
>0 $). With respect to the TF great controversy still exists
upon its direction \cite{ior,so,pita} or even more, upon its
proper existence \cite{tho,vol}.

In the present work,
the reservoir vector operator $\bf B$ of Eq. (\ref{hin}) is
chosen as to describe qp scattering by the vortex. In
terms of creation $a^+_{\bf{q}}$ and annihilation $a_{\bf{q}} $
operators for a qp of momentum $\bf{q}$ it reads,
\begin{equation}
 {\bf B} =  \sum_{ {\bf k} , {\bf q} }  \, \,   ( {\bf k} -  {\bf q} ) 
  \,\,  \Lambda_{k,q}    \,\,
a_{{\bf k}}^+ \,  a_{ {\bf q}} \,\, ,
\label{bintn}
\end{equation}
being $ \bf{k} -\bf{q} $ the transferred momentum and 
$ \Lambda_{k,q}$ a vortex-qp coupling constant which, in order to 
preserve isotropy,
 is assumed to depend only on the modulus of qp
momentum ($q=|{\bf q}|$).

To calculate the drag coefficients we must first
compute the correlation function in (\ref{corr}). Taking into
account the time evolution $a_{ \bf{q}}(t)=  a_{ \bf{q}}(0)
e^{-i w_q t }$ and after some calculations, we have
\begin{equation}
< B_j(t)\, B_j> =\frac{1}{2} \sum_{ \bf{k} ,  \bf{q} } 
|\Lambda_{k,q}|^2 
(k^2+q^2) e^{i(w_k-w_q) t}  n(w_k) [1 + n(w_q)]    \,\, ,
\label{bx}
\end{equation}
where $n(w_q)$ denotes the mean number of qp's of energy $\hbar w_q$,
\begin{equation}
n (w_q) = <a_{\bf{q}}^+ \,  a_{ \bf{q}}>  = 
\frac{1}{e^{\beta \hbar w_q} -1 }  \, \, .
\label{eneq}
\end{equation}

Note that at temperature $T=0$ the correlation function
(\ref{bx}) vanishes due to the  factor $n(w_k)$, yielding a
vanishing drag force  as expected.  This behavior would not be
reproduced by a linear interaction in the qp operators, as
for example the one proposed 
in the Hamiltonian of Ref. \cite{niu}.

The parameter $\gamma$  (Eq. (\ref{integ})) can be written in terms of 
the Fourier transform $\psi[w]$ of the imaginary
part of the correlation function (\ref{bx}) as
\begin{eqnarray}
{\cal R}e (\gamma)  & = & \frac{4}{i\hbar}\Omega\,\,\,
 {\cal P} \int_{0}^{\infty}\,dw 
\frac{w}{ \Omega^2-w^2 } \psi[w]
\label{rein} \\
{\cal I}m (\gamma) & = & \frac{ 2\pi}{i\hbar}\Omega \psi[\Omega]
\label{imin}
\end{eqnarray}
where ${\cal P}$ refers to the principal part and 
\begin{equation}
\psi[w] = 
\frac{i}{4}  \sum_{ {\bf k},  {\bf q} } (k^2+q^2)
 |\Lambda_{k,q}|^2 [n(w_q)-n(w_k)]\delta (w_k-w_q-w)\,\, .
\label{delsin}
\end{equation}
The above expressions (\ref{imin}) and (\ref{delsin}) for the 
dissipative
coefficient in (\ref{flo}) can be interpreted from the scattering 
processes
embodied in our interaction Hamiltonian. In fact, we first note that 
the vortex Hamiltonian (\ref{Hfree}) has an equally spaced level 
spectrum of
separation $\hbar\Omega$ (the so-called Landau levels\cite{landau}) 
and,
in addition, the operator ${\bf v}$ which couples the vortex to the 
reservoir
qp's, can be expressed from (\ref{vfree}) as a linear combination of 
creation
and destruction operators of a vortex energy quantum\cite{landau}.
Thus, according to (\ref{hin}) and (\ref{bintn}), we may identify the 
scattering processes embodied in our model as those involving one 
vortex
energy quantum $\hbar\Omega$ jointly with the creation and destruction 
of one qp. Such processes can also be identified from (\ref{delsin}) as 
follows.
Each term proportional to $n(w_q)$ represents the process by which a qp
of energy $\hbar w_q$ combines with a vortex quantum $\hbar\Omega$ to 
form
a qp of energy $\hbar w_k=\hbar w_q+\hbar\Omega$ (this may be seen from
(\ref{imin}) and the Dirac delta in (\ref{delsin})). The weight 
$n(w_q)$ of
this process can be easily understood by taking into account the {\em
thermal} origin of the ``incoming'' qp of energy
$\hbar w_q$ in contrast to the
{\em interaction} origin of the ``outgoing'' qp of energy $\hbar w_k$.
Similarly, the terms weighted by $n(w_k)$ in (\ref{delsin}) represent 
the
annihilation of a qp of energy $\hbar w_k$ jointly with the creation of 
a vortex quantum $\hbar\Omega$ and a qp of energy $\hbar w_q$. Finally,
 we note that the positive sign of the dissipative coefficient 
${\cal I}m (\gamma)$ arises from the factor $n(w_q)-n(w_k)>0$ in 
(\ref{delsin})
i. e., the weight of processes involving vortex energy loss must be 
greater
than those causing vortex energy gain.

At low enough temperatures it is to be expected that only scattering 
processes
involving phonons should be relevant. Therefore, we shall use the 
phonon dispersion 
relation $w_q=c_sq$ (being
$c_s$ the sound velocity) and impose
a cutoff momentum to the qp's. 
This amounts to neglect all the scattering processes
which yield not a phonon as the ``outgoing'' qp. Note that the 
``incoming''
{\em thermal} qp will be surely a phonon. Under such an approximation
 Eq. (\ref{delsin}) can be written as,

\begin{eqnarray}
\psi[w] & =  & \frac{i}{4c_s^2}
\int_{0}^{\infty}\,dw'\, S(w',w'+w) [w'^2 +(w'+w)^2]\nonumber\\
 & \times & [n(w') - n(w'+w)] 
\label{psiwm}
\end{eqnarray}
where we introduce the so-called scattering function \cite{mar}, 
defined as
\begin{equation}
S(w',w'') = \sum_{{\bf k},{\bf q} } |\Lambda_{k,q}|^2 
\delta (w'-w_q)\delta(w''-w_k)  \,\,
\label{scat}
\end{equation}
and being related to the scattering of the environmental 
excitations between states of frequencies $w'$ and $w''$.

 The integral in
(\ref{psiwm}) has to be solved numerically except in the limit $
T \rightarrow 0 $. In fact, for $\hbar w/k_BT\rightarrow\infty$  the
gain term $n(w'+w)$ can be neglected and the loss one $n(w')$ 
``filters''
all but the lowest frequency ``incoming'' phonons. This means that the 
factors 
accompanying $n(w')$ in (\ref{psiwm}) can be approximated to lowest 
order
in $w'$. In particular,  the scattering function (\ref{scat}) is 
assumed
to be a continuous symmetric function of both variables \cite{mar} 
satisfying $S(w',w)\simeq S(w)\,w'^p$ for
$w'\rightarrow 0$. Thus equation (\ref{psiwm}) can be approximated
for $T\rightarrow 0$ as follows,
\begin{eqnarray}
\psi[w] & \simeq & \frac{i}{4c_s^2}w^2\,S(w)\,\int_0^{\infty}
dw'\,w'^p\,n(w')=\nonumber\\
 & = & \frac{i}{4 c_s^2}
\, p!\,\zeta(p+1)\,w^2\,S(w) 
\left(\frac{k_BT}{\hbar}\right)^{p+1} 
\,\, ,
\label{Ttend0}
\end{eqnarray}
where $\zeta(n)\,\,\,(n\geq 2)$ denotes the Riemann zeta
 function. 

It is convenient to notice that our model is unable to provide an
{\em a priori} explicit form for the scattering function, because
we treat vortex and qp excitations as separate entities which are 
assumed
to interact by means of a generic Hamiltonian. Any additional 
information
should be based upon experimental results or more fundamental theories.
In fact, for the lowest temperature domain, only theoretical results 
are 
at present available and they predict a $T^5$ dependence for the LF 
\cite{ior}.
Hence, we set $p=4$ in (\ref{Ttend0}) and accordingly a low frequency 
$\sim
w^4$ behavior for the scattering function. To perform numerical 
calculations
which illustrate our results we shall assume a simple form for the 
scattering function (\ref{scat}), as
 a generalization of the usual super-Ohmic dissipation with an 
exponential cutoff \cite{weiss}:
\begin{equation}
S(w',w'') \propto
 w'^4 e^{- w'/w_o} w''^4 e^{- w''/w_o} \,\, ,
\label{sca2}
\end{equation}
being $w_o$ a frequency cutoff parameter
 which allows us to select only the phonon
part of the $^4$He qp excitations spectrum i. e., 
the frequencies below $w_c\simeq$1.2 ps$^{-1}$. This is evident from
 Fig. 1 where we plot the one variable scattering function $S(w,w)$
for $w_o$=0.06 ps$^{-1}$. In addition, 
we plot the frequency spectrum  $f(w)$ of
the normal fluid density (i. e., $\rho_n=\int_0^\infty
f(w)dw$) for T=0.4 K, which shows that at most the first
half of the phonon spectrum makes a relevant contribution to the
qp excitations according to a similar behavior for
the one variable scattering function.

From Eqs. (\ref{rein}) and (\ref{imin}) we note that the value
of cyclotron frequency, which in turn depends on the vortex mass
$m_v$, (see Eq. (\ref{Ome})) is necessary for the determination
of the drag force.  Unfortunately, there is also a
controversy regarding the calculation of the vortex
mass\cite{niu,duan2,duan,niu2} that makes the possible value of
$\Omega$ to range from a lower bound\cite{duan2}
 $\Omega_{min}\simeq 0.1$
ps$^{-1}$  to the upper bound\cite{do} $ \Omega_{max}\simeq
3$ ps$^{-1}$.  This suggests that a study of the
drag force dependence on $\Omega$ could be illustrative.

In Fig. 2  we show the coefficients of both,  the transverse and
the longitudinal components of the force, that is ${\cal R}e
(\gamma)$ and ${\cal I}m (\gamma)$ respectively, as functions of $
\Omega $ for different temperatures.  We see that, apart from a
change of scale, the shape of curves does not change appreciably
along the temperature range $0<T\leq 0.4K$  that we are considering.
The only effect seems to be a shift to low frequencies
with increasing temperatures (note that for a fixed vortex mass,
the cyclotron frequency (\ref{Ome}) decrease with increasing
temperature due to the factor $\rho_s$, however this variation
is negligible for  $T<0.4K$). 

Regarding  ${\cal I}m
(\gamma)$ we first indicate that according to (\ref{imin}), the
plots of this function in Fig. 2 turn out to be proportional to
the Fourier transform of $\dot{\psi}(t)$.
In addition, we notice that ${\cal I}m(\gamma)$ vanishes for
$\Omega> w_c $.  Such absence of dissipation is simply
understood as a direct consequence of energy conservation in the
scattering processes. This is imposed by the Dirac delta in 
(\ref{delsin}),
since it reflects the fact that  phonon-to-phonon scattering events can
only take place if the vortex energy quantum does not exceed the most
energetic phonon of the spectrum.
 Therefore, we may
conclude that dissipation due to phonon$\rightarrow$phonon
scattering should be negligible unless
 the vortex mass yields a value of cyclotron
frequency less than $ w_c \simeq$1.2 ps$^{-1}$, which is an 
intermediate value
between the quoted frequencies $\Omega_{min}$ and
$\Omega_{max}$. Moreover, the value $\hbar\Omega_{max}$, arising from a
usual hydrodynamical prescription for the calculation of the vortex 
mass\cite{do}, exceeds the energy of any
undamped qp excitation\cite{pist}. So, we may extend our previous 
conclusion
by saying that any dissipation via qp$\rightarrow$qp scattering events
 should
be negligible for such value of vortex mass, suggesting thus a possible
scenario of dissipation via multi-qp scattering processes\cite{marko}.

Regarding the TF, we remark that if it were possible to draw
some experimental information about at least the direction of
this component, this would also shed more light on the value of
the vortex mass. In fact, from Fig. 2 we note that at
intermediate values between $\Omega_{min}$ and $w_c$, the TF
changes its direction. In particular, a positive (negative) value
of this component implies   $ \Omega > 0.45 $ ps$^{-1}$ ($\Omega < 
0.3$ ps$^{-1}$), 
while a change of sign yields some intermediate value of
$\Omega$.  Unfortunately no experimental data are available for
temperatures below $ 0.4 $ K.

It is important to note, in concluding this report, that our study
has been restricted to a strictly rectilinear vortex, neglecting thus
any possible contribution to the forces arising from vortex line
curvatures. Clasically, vortex lines can be deformed as helical waves
known as Kelvin waves, and there is experimental evidence of similar
modes for quantized vortices in Helium II \cite{do}. From the 
theoretical
viewpoint, such waves could be regarded as a particular form of 
elementary
excitation bound to the vortex. In fact, it has been shown that the 
elementary
excitation spectrum of an imperfect Bose gas in presence of a straight 
vortex
line consists of both, phonons and Kelvin-like waves \cite{pife}.
Viewed from this perspective, being two different kinds of independent 
elementary excitations, the consequence of scattering processes 
involving
phonons and Kelvin modes appears eventually as a possible secondary 
effect upon the values of longitudinal and transverse forces.
This issue has been scarcely addressed in the literature. In 
particular,
Sonin  has long ago performed a simplified study \cite{sofi} neglecting
the vortex velocity due to Kelvin modes, but taking into account the
effect of such modes on the scattering of phonons by means of an 
average over a classical Rayleigh-Jeans distribution for the 
oscillations of the vortex filament. He concluded that such effect
could at most modify the calculated forces for a strictly rectilinear
vortex in 1.7 \% at $T=0.5K$. Unfortunately, within our formalism
a quantitative study 
 of the dynamics of a true three-dimensional
curved vortex filament, exhibits a high degree of difficulty.

In summary, starting from a microscopic model we have 
performed a calculation of
the drag force on a moving vortex due to the scattering of
qp excitations at temperatures below 0.4 K.  We
have also discussed, by analyzing both the longitudinal and 
transverse forces as functions
of the cyclotron frequency, some important
implications  in the determination of
possible scattering processes leading to dissipation, according to the
values of vortex mass found in the literature.

\ack

This work was supported by grant PEI 0067/97 from 
Consejo Nacional de Investigaciones Cient\'{\i}ficas y
T\'ecnicas, Argentina.


\newpage

\begin{figure}
\vspace{2cm}
        \begin{center}
\epsfxsize=11cm
        \epsfbox{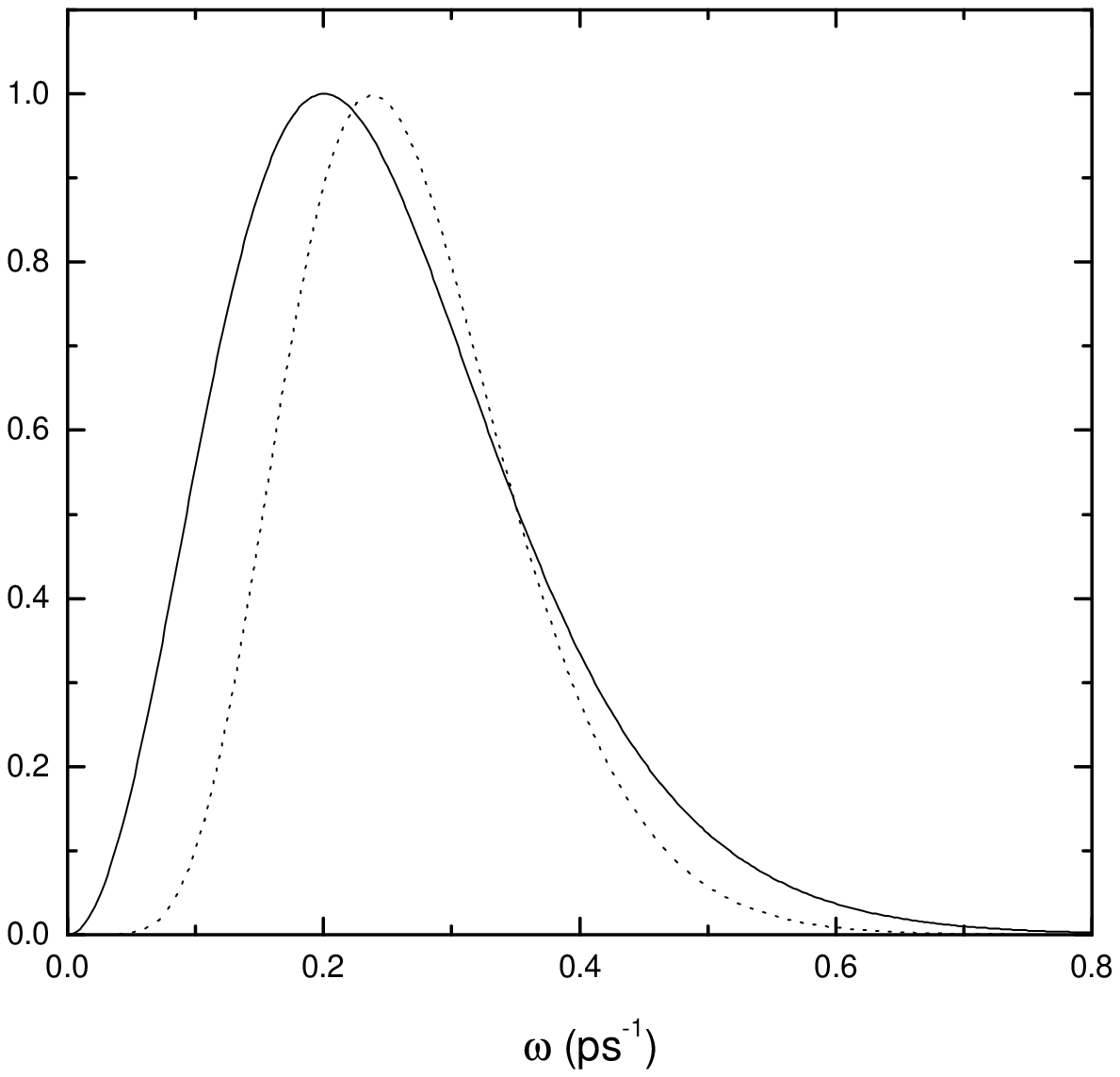}  
        \end{center}
        \caption{ \label{fig1}
                One variable scattering function,
 $ S(w,w) $ 
for $w_o$=0.06 ps$^{-1}$ (dotted line) and frequency spectrum  
$f(w)$ of the 
normal fluid density  for T=0.4 K (solid line). Different scales were
 used
for each function in order to normalize the size of both peaks.}

\end{figure}

\newpage
\begin{figure}
\vspace{2cm}
        \begin{center}
\epsfxsize=11cm
        \epsfbox{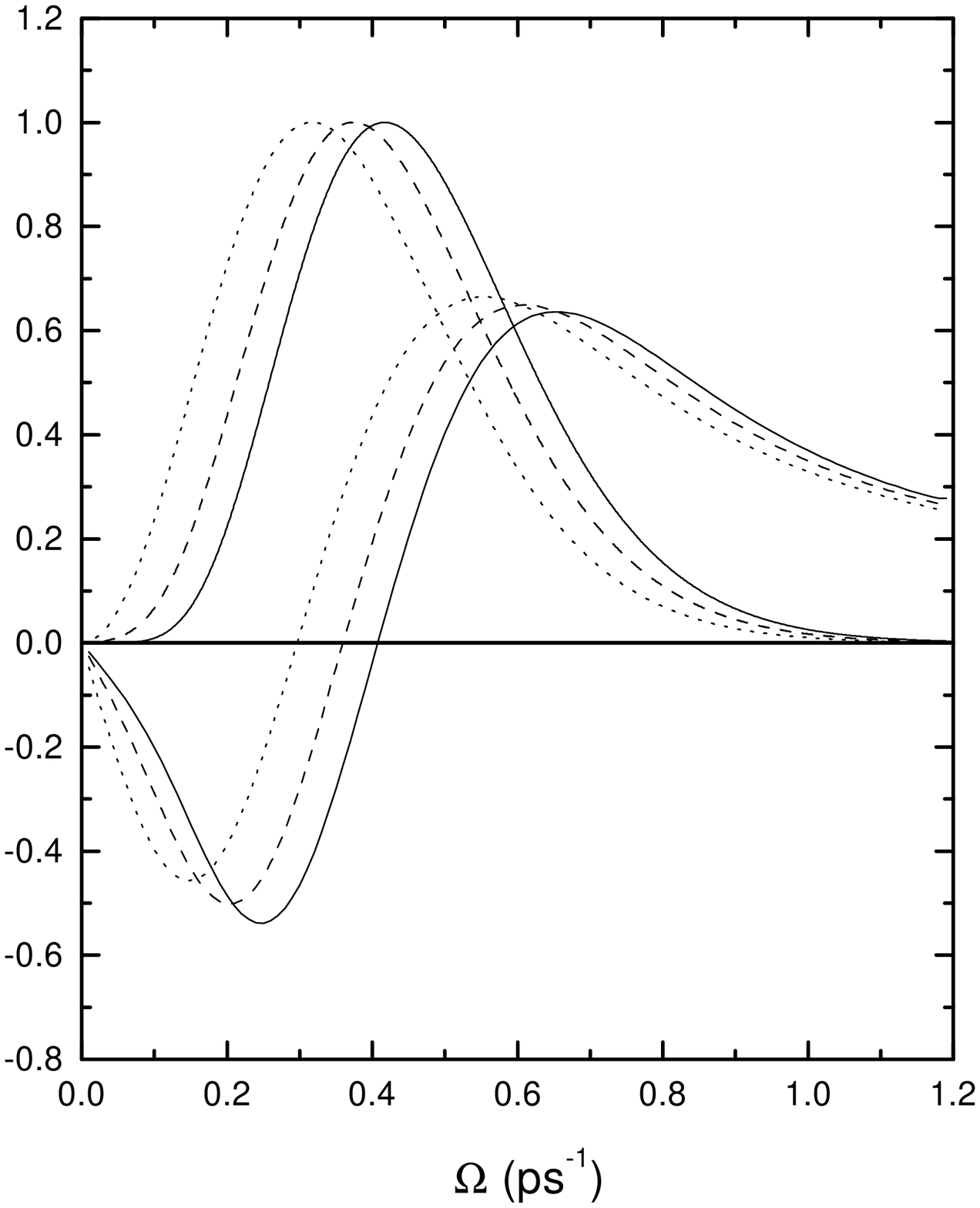}  
        \end{center}
        \caption{ \label{fig2}
                Coefficients of the TF
(${\cal R}e (\gamma)$) and the LF (${\cal I}m (\gamma)$) for 
$T\rightarrow 0$
(Eq. (20)) (solid line); $T=0.1K$ (dashed line) and
$T=0.4K$ (dotted line). Different scales were used for each temperature
in order to normalize the size of the ${\cal I}m (\gamma)$ peak. The 
relative
scale factors are 105.82 and infinity for $T=0.4K$/$T=0.1K$ and
$T=0.1K$/$T\rightarrow 0$, respectively. The exponential parameter
$w_o$=0.06 ps$^{-1}$
was used in Eq. (21).}

\end{figure}

\end{document}